\documentclass{emulateapj}

\shorttitle{Effects of Leakage Neutral Particles on Shocks}

\shortauthors{Yutaka Ohira}

\begin{document}

\title{Effects of Leakage Neutral Particles on Shocks}

\author{Yutaka Ohira}

\begin{abstract}
In this paper we investigate effects of neutral particles on 
shocks propagating into the partially ionized medium.
We find that 
for $120~{\rm km~s}^{-1}<u_{\rm sh}<3000~{\rm km~s}^{-1}$ 
($u_{\rm sh}$ is the shock velocity), 
about ten percent of upstream neutral particles 
leak into the upstream region from the downstream region.
Moreover, we investigate how the leakage neutral particles affect the upstream 
structure of the shock and particle accelerations. 
Using four fluid approximations 
(upstream ions, upstream neutral particles, leakage neutral particles and pickup ions), 
we provide analytical solutions of the precursor structure due to leakage neutral particles.
It is shown that the upstream flow is decelerated in the precursor region and 
the shock compression ratio becomes smaller than without leakage neutral particles, 
but the total compression ratio does not change. 
Even if leakage of neutral particles is small (a few percents of total upstream particles), 
this smaller compression ratio of the shock can explain steep gamma-ray spectra 
from young supernova remnants. 
Furthermore, leakage neutral particles could amplify the magnetic field 
and heat the upstream region. 
\end{abstract}


\affil{Department of Physics and Mathematics, Aoyama Gakuin University, 
5-10-1 Fuchinobe, Sagamihara 252-5258, Japan; ohira@phys.aoyama.ac.jp}
\section{Introduction}

Supernova remnants (SNRs) are thought to be the origin of Galactic 
cosmic rays (CRs). 
The most popular SNR acceleration mechanism is diffusive shock 
acceleration (DSA) \citep{axford77,krymsky77,bell78,blandford78}. 
In fact, {\it Fermi\/} and {\it AGILE\/} observed GeV gamma rays due 
to CRs from SNRs \citep[e.g.][]{abdo09,abdo10,tavani10,giuliani11, giordano12}.
However, gamma-ray spectra from SNRs are steeper than that 
expected from the standard DSA theory. 
The steep spectra can be interpreted as effects of energy-dependent 
escape \citep{ptuskin05,ohiraetal10,caprioli10} and diffusion 
\citep[e.g.][]{ohiraetal11} for middle-aged SNRs ($\sim10^4~{\rm yrs~old}$). 
For young SNRs ($\lesssim10^3~{\rm yrs~old}$), some ideas to explain the 
steep spectra have been proposed 
\citep{kirk96,zirakashvili09,ohiraetal09,ohira10,bell12} but it is still an open issue.

On the other hand, some authors considered effects of neutral particles (such as hydrogen atoms)
on particle accelerations and shock structures. 
SNR shocks propagating into a partially ionized medium have been 
observed as Balmer-dominated shocks \citep{chevalier78}. 
Moreover, X-ray synchrotron radiation has been observed from the 
Balmer-dominated shocks \citep{cassam08,helder09}. 
A neutral fraction of the interstellar medium around SNRs is often found to be 
order of unity \citep{gha00, gha02}. 
Neutral particles reduce growth rates of CR streaming instabilities 
which are indispensable for DSA \citep{drury96, reville07}. 
In contrast, ions produced from the neutral particles trigger other plasma instabilities 
and are important for the injection into particle accelerations \citep{ohiraetal09,ohira10}
A recent review of Balmer-dominated shocks can be found in \citet{heng10}.

Interactions between neutral particles and ions have been directly 
observed in the solar wind. 
There are two peculiar particles, energetic neutral atoms \citep{mccomas09} 
and pickup ions \citep{gloeckler93} in the solar wind.
Although their origin has not been completely understood, 
an attractive idea has been proposed. 
Neutral particles penetrate into the inner solar system from 
the surrounding local interstellar medium.
The neutral particles have a drift velocity comparable to the solar wind velocity 
in the rest frame of the solar wind. 
After they are ionized by charge exchange and photo ionization, 
they gyrate around magnetic field lines of the solar wind and 
their mean velocity becomes the solar wind velocity in the observer frame, 
so that they become pickup ions with a large velocity dispersion in the solar wind. 
After passing over the termination shock, some pickup ions become neutral atoms 
by charge exchange. 
The neutral atoms can propagate toward the sun and are observed as energetic 
neutral atoms.

Applying these pictures to SNR shocks propagating into a partially ionized 
medium, we expect leakage of neutral particles from the downstream 
region to the upstream region. 
\citet{raymond08} discussed formation and neutralization of pickup ions 
in the downstream region of SNR shocks. 
Therefore, we can expect leakage neutral particles not only from 
pickup ions produced in the upstream region but also from 
pickup ions produced in the downstream region. 
Upstream pickup ions originate from leakage neutral particles, 
so that downstream pickup ions should be the dominant source of 
leakage neutral particles.

The leakage neutral particles become pickup ions by 
collisional ionization or charge exchange in the upstream region. 
If leakage is significant, the upstream flow is decelerated and heated 
by the pickup ions.
As the result, a precursor is produced by leakage neutral particles.  
Very recently, \citet{blasi12} showed that neutral particles really leak into the upstream 
region from the downstream region by solving the Vlasov equation 
of neutral particles. 
They showed the formation of the precursor due to the leakage neutral particles 
and provided the precise velocity distribution of neutral particles 
by assuming that only the ion distribution is a Maxwellian.

In this paper, we investigate the precursor structure 
by a different approach which is a four fluid approximation. 
We consider upstream ions, upstream neutral particles, 
leakage neutral particles and pickup ions, respectively. 
Then, we obtain analytical solutions of the precursor structure.
Our results in this paper are qualitatively similar to that of \citet{blasi12}.

We first estimate the number density and the velocity of leakage 
neutral particles in Section 2.
We then provide some lengthscales for collisional ionization 
and charge exchange in Section 3, 
and provide four fluid models to describe the precursor structure in Section 4, 
and provide two approximate solutions in Sections 4.1 and 4.2. 
Section 5 is devoted to the discussion. 
\section{Distribution Function of Leakage Neutral Particles}
\label{sec:2} 
In this section, we estimate the distribution function of leakage neutral particles 
in the shock rest frame. 
Leakage neutral particles originate from hot neutral particles produced 
in the downstream region. 
We here consider charge exchange, collisional ionization and Coulomb collision 
as interactions in the downstream region. 
Although ionization by electrons could be important, 
it depends on the electron temperature which has not been understood 
yet \citep{ohira07,ohira08,rakowski08}.
We here do not take into account ionization by electrons because 
it is sub-dominant compared to that by protons 
as long as the relative velocity is larger than about $2000~{\rm km~s}^{-1}$ 
\citep[e.g.][]{heng07} which is a typical shock velocity of young SNRs.
Cross sections of charge exchange and collisional ionization depend on 
a relative velocity, $u_{\rm rel}$ \citep[e.g.][]{heng07}.
Collisional ionization of hydrogen atoms by protons is dominant 
for $u_{\rm rel}\gtrsim 3000~{\rm km~s^{-1}}$ and its cross section is typically 
$\sigma_{\rm i} \sim 10^{-16}~{\rm cm^2}$.
On the other hand, charge exchange is dominant for 
$u_{\rm rel}\lesssim 3000~{\rm km~s^{-1}}$ 
and its cross section between protons and hydrogen atoms is 
about $\sigma_{\rm ce}\sim 10^{-16}~{\rm cm^2}$ 
for $u_{\rm rel} \sim 3000~{\rm km~s^{-1}}$ 
and $\sigma_{\rm ce}\sim 10^{-15}~{\rm cm^2}$ 
for $u_{\rm rel}\lesssim 1000~{\rm km~s^{-1}}$.
The ionization cross section of hydrogen atoms by hydrogen atoms, 
$\sigma_{\rm i,HH}$, (${\rm H+H\rightarrow p+e^-+H}$) is quite similar 
to that by protons \citep{barnett90}, 
so that we let $\sigma_{\rm i,HH} =\sigma_{\rm i}$.
All relative velocities in the downstream region are typically the shock velocity, 
$u_{\rm sh}$, which is about $3000~{\rm km~s}^{-1}$ for young SNRs, 
so that both collisional ionization and charge exchange are important processes 
for neutral particles.

A neutral fraction of the interstellar medium around SNRs is often found to be 
order of unity \citep{gha00, gha02}. 
The typical ISM density is $1~{\rm cm}^{-3}$.
Therefore, we use $n_{\rm n} = n_{\rm ion} = 0.5~{\rm cm}^{-3}$ 
as fiducial values in this paper, where  $n_{\rm n}$ and $n_{\rm ion}$ 
are number densities of upstream neutral particles and ions, respectively.

In the shock rest frame, upstream ions are decelerated 
and heated at the shock, while upstream neutral particles 
are not decelerated because the shock dissipation is due to 
electromagnetic interactions \citep{chevalier78}.
After passing over the shock front, the upstream neutral particles 
are mainly ionized by hot ions in the downstream region. 
Then, the upstream neutral particles become 
pickup ions in the downstream region \citep{raymond08}. 
Therefore, the ionization timescale of upstream cold neutral 
particles in the downstream region, $t_{\rm i,cold}$, is given by
\begin{equation}
t_{\rm i,cold} = \frac{1}{\left(\sigma_{\rm i}+\sigma_{\rm ce}\right)n_{\rm ion,hot}u_{\rm rel}} ~~,
\end{equation}
where $n_{\rm ion,hot}=rn_{\rm ion}$ is the number density of hot ions 
in the downstream region and $r$ is the shock compression ratio. 
Hot neutral particles are produced from hot ions by charge exchange 
in the ionization lengthscale of penetrating neutral particles, 
$L_{\rm ion,down}=u_{\rm sh} t_{\rm i,cold}$, where $u_{\rm sh}$ is the shock velocity. 
The charge exchange timescale of hot ions, $t_{\rm ce}$, is given by 
\begin{equation}
t_{\rm ce} = \frac{1}{\sigma_{\rm ce}n_{\rm n}u_{\rm rel}} ~~.
\label{eq:tce}
\end{equation}
The crossing timescale of hot neutral particles that move toward the 
shock with a negative velocity of $v_x$, is given by
\begin{equation}
t_{\rm cross,n} (v_x)=\frac{L_{\rm ion,down}}{|v_x|}=t_{\rm i,cold}\frac{u_{\rm sh}}{|v_x|}~~,
\end{equation}
where $v_x$ is a velocity in the direction of the shock normal and $v_x<0$ 
($x=0$ and $x=-\infty$ are the positions of the shock and the far upstream region, respectively.). 
Hot neutral particles are mainly ionized by hot ions, so that the ionization 
timescale of hot neutral particles, $t_{\rm i,hot}$, is given by 
\begin{equation}
t_{\rm i,hot} = \frac{1}{\left(\sigma_{\rm i}+\sigma_{\rm ce}\right)n_{\rm ion,hot}u_{\rm rel}}~~.
\label{eq:tihot}
\end{equation}
Therefore, in the shock rest frame, 
the steady-state distribution function of hot neutral particles 
with $v_x<0$ at the shock, that is, the distribution function of leakage 
neutral particles at the shock is approximately given by
\begin{equation}
f_{\rm leak,sh}(\mbox{\boldmath $v$}) = \frac{f_{\rm ion,hot}(\mbox{\boldmath $v$})}{t_{\rm ce}}\times \min(t_{\rm cross,n}(v_x),t_{\rm i,hot}) ~~,
\label{eq:fleaksh}
\end{equation}
where $ f_{\rm ion,hot}(\mbox{\boldmath $v$})$ is 
the distribution function of downstream hot ions in the shock rest frame.

Although the velocity distribution of hot ions in the downstream region, 
$f_{\rm ion,hot}(\mbox{\boldmath $v$})$, 
has not been understood for SNR shocks, especially, 
in the partially ionized medium, 
we here assume that the dominant source of leakage neutral particles is 
pickup ions produced in the downstream region because 
they should have the large velocity dispersion and the high density 
compared to those of other components. 
The velocity distribution of pickup ions produced in the downstream region 
becomes approximately an isotropic shell distribution in the downstream rest 
frame \citep{raymond08}.
The isotropic velocity in the downstream rest frame, 
$v_{\rm d}$, is the velocity of upstream neutral particles 
in the downstream rest frame, $(1-r^{-1})u_{\rm sh}$, 
where $r$ is the shock compression ratio. 
Therefore, the ionization timescale of hot neutral particles is smaller than 
the crossing timescale of hot neutral particles in the ionization region 
($t_{\rm i,hot}<t_{\rm cross,n}(v_x)$) because $v_x \geq -(1-2r^{-1})u_{\rm sh}$ 
for pickup ions produced in the downstream region. 
In this case, leakage neutral particles with the negative velocity, $v_x$, 
are produced in the production region, $0\leq x\leq t_{\rm i,hot}|v_x|$. 
Then, the crossing timescale of downstream pickup ions in the production region is given by
\begin{equation}
t_{\rm cross,PUI}(v_x)= r\frac{|v_x|}{u_{\rm sh}}t_{\rm i,hot}~~. 
\end{equation}
Furthermore, from Equation~(\ref{eq:fleaksh}), the distribution function of leakage neutral particles 
in the shock rest frame, $f_{\rm leak,sh}(\mbox{\boldmath $v$})$, is expressed by
\begin{equation}
f_{\rm leak,sh}(\mbox{\boldmath $v$}) = \frac{t_{\rm i,hot}}{t_{\rm ce}}\frac{n_{\rm PUI,down}}{4\pi v_{\rm d}^2} \delta \left [v_{\rm d}-\left (1-r^{-1}\right)u_{\rm sh}\right]~~,
\label{eq:fleaksh2}
\end{equation}
where $v_{\rm d}=\{(v_x-u_{\rm sh}r^{-1} )^2+v_y^2+v_z^2 \}^{1/2}$ is 
a particle speed in the downstream rest frame and 
$v_y, v_z, \delta [...]$ and $n_{\rm PUI,down}$ are velocities perpendicular to the shock normal,  
the delta function and the number density of downstream pickup ions 
contributing to the leakage neutral particles, respectively.

Downstream pickup ions are eliminated by charge exchange and Coulomb collision. 
The relaxation timescale of pickup ions by Coulomb collision is given by \citep{spitzer62}
\begin{equation}
t_{\rm Coulomb} = \frac{m_{\rm p}^2u_{\rm rel}^3}{8\pi n_{\rm hot,ion}e^4 \ln \Lambda}~~,
\label{eq:tcoulomb}
\end{equation}
where $e$ and $\ln \Lambda$ are the elementary charge and the Coulomb logarithm, respectively.
Then, the number density of pickup ions contributing to the leakage neutral particles, 
$n_{\rm PUI,down}$, is approximately given by 
\begin{equation}
n_{\rm PUI,down} = \frac{n_{\rm n}}{t_{\rm i,cold}} \times \min \left(t_{\rm cross,PUI}(v_x),t_{\rm ce},t_{\rm Coulomb} \right)~~.
\label{eq:npuidown}
\end{equation}
The crossing timescale of downstream pickup ions, $t_{\rm cross,PUI}(v_x)$, is 
always smaller than the charge exchange timescale of hot ions, $t_{\rm ce}$, 
as long as $n_{\rm ion}/n_{\rm n}\geq 0.5$. 
We here consider the case of $t_{\rm cross,PUI}<t_{\rm ce}$ 
because the shock dissipation has not been completely understood yet 
for the low ionization fraction. 
The maximum value of $t_{\rm cross,PUI}(v_x)$ is about $t_{\rm i,hot}$. 
Therefore, we consider two cases: $t_{\rm i,hot}<t_{\rm Coulomb}$ 
and $t_{\rm i,hot}>t_{\rm Coulomb}$. 
From the condition, $t_{\rm i,hot}=t_{\rm Coulomb}$, we obtain the critical shock velocity, 
$u_{\rm sh,c}$, as 
\begin{eqnarray}
u_{\rm sh,c} &=& \left\{\frac{8\pi e^4 \ln \Lambda}{m_{\rm p}^2\left(\sigma_{\rm i}+\sigma_{\rm ce}\right)} \right \}^{\frac{1}{4}} \nonumber \\
&=& 120~{\rm km~s}^{-1} \left( \frac{\ln \Lambda}{40} \right)^{\frac{1}{4}} \left( \frac{\sigma_{\rm i}+\sigma_{\rm ce}}{10^{-15}~{\rm cm}^2} \right)^{-\frac{1}{4}}~~,
\end{eqnarray}
where we assume $u_{\rm rel}\sim u_{\rm sh}$.

For $t_{\rm i,hot} < t_{\rm Coulomb}$ ($u_{\rm sh} > u_{\rm sh,c}$), 
relaxation by Coulomb collision is negligible in the production region of leakage neutral particles, 
$0\leq x\leq t_{\rm i,hot}|v_x|$. 
Then, from Equations~(\ref{eq:tce}), (\ref{eq:tihot}), (\ref{eq:fleaksh2}) and (\ref{eq:npuidown}), 
the distribution function of leakage neutral particles in the shock rest frame, 
$f_{\rm leak,sh}(\mbox{\boldmath $v$})$, is expressed by
\begin{eqnarray}
f_{\rm leak,sh}(\mbox{\boldmath $v$}) &=& rn_{\rm n} \frac{n_{\rm n}}{n_{\rm hot,ion}} \frac{\sigma_{\rm ce}}{\sigma_{\rm i}+\sigma_{\rm ce}} \frac{|v_x|}{u_{\rm sh}} \nonumber \\
&&\times \frac{1}{4\pi v_{\rm d}^2} \delta \left [v_{\rm d}-\left (1-r^{-1}\right)u_{\rm sh}\right]~~,
\label{eq:fleaksh3}
\end{eqnarray}
where $v_{\rm d}=\{(v_x-u_{\rm sh}r^{-1} )^2+v_y^2+v_z^2 \}^{1/2}$.
Then, the number density of leakage neutral particles at the shock, $n_{\rm leak,sh}$ is given by 
\begin{eqnarray}
n_{\rm leak,sh} &=& \int_{-\infty}^{\infty}{\rm d}v_y\int_{-\infty}^{\infty} {\rm d}v_z \int_{-\infty}^{0} {\rm d}v_x f_{\rm leak,sh} (\mbox{\boldmath $v$}) \nonumber \\
&=& \frac{1}{12}n_{\rm n} \frac{n_{\rm n}}{n_{\rm ion}} \frac{\sigma_{\rm ce}}{\sigma_{\rm i}+\sigma_{\rm ce}} F(r)~~, \label{eq:nleaksh}
\end{eqnarray}
where $F(r)=3(1-2r^{-1})^2/(1-r^{-1})$ ($F=1$ for $r=4$).
Hence, for $120~{\rm km~s}^{-1}<u_{\rm sh}\lesssim 3000~{\rm km~s}^{-1}$, 
about ten percent of upstream neutral particles leak 
into the upstream region from the downstream region.
For $u_{\rm sh}> 3000~{\rm km~s}^{-1}$, 
the cross section of ionization is larger than that of charge exchange 
($\sigma_{\rm i} > \sigma_{\rm ce}$), 
so that leakage of neutral particles becomes smaller. 
Moreover, in the shock rest frame, the mean velocity of leakage neutral particles at the shock, $u_{\rm leak,sh}$, is given by 
\begin{eqnarray}
u_{\rm leak,sh} &=& \frac{1}{n_{\rm leak,sh}}\int_{-\infty}^{\infty}{\rm d}v_y\int_{-\infty}^{\infty} {\rm d}v_z \int_{-\infty}^{0} {\rm d}v_x  f_{\rm leak,sh} (\mbox{\boldmath $v$})v_x \nonumber \\
&=&-\frac{1}{3} u_{\rm sh} G(r) ~~.
\label{eq:uleaksh}
\end{eqnarray}
where $G(r)=2(1-2r^{-1})$ ($G=1$ for $r=4$).
Hence, the mean velocity of leakage neutral particles in the shock rest frame is 
about one third of the shock velocity for $u_{\rm sh}> u_{\rm sh,c}$. 
For the smaller compression ratio, $n_{\rm leak,sh}$ and $|u_{\rm leak,sh}|$ become smaller 
and there is no leakage for $r\leq 2$.

For $t_{\rm i,hot} > t_{\rm Coulomb}$ ($u_{\rm sh} < u_{\rm sh,c}$), 
relaxation by Coulomb collision is not negligible. 
For leakage neutral particles satisfying $t_{\rm Coulomb}\leq t_{\rm cross,PUI}(v_x)$ 
($v_x\leq -r^{-1}u_{\rm sh}(u_{\rm sh}/u_{\rm sh,c})^4$), 
from Equations~(\ref{eq:tce}), (\ref{eq:fleaksh2}), (\ref{eq:tcoulomb}) and (\ref{eq:npuidown}), 
the distribution function of leakage neutral particles at the shock, $f_{\rm leak,sh}(\mbox{\boldmath $v$})$, is expressed by
\begin{eqnarray}
f_{\rm leak,sh} (\mbox{\boldmath $v$})&=&\frac{1}{r}n_{\rm n}\frac{n_{\rm n}}{n_{\rm ion}}\frac{\sigma_{\rm ce}m_{\rm p}^2u_{\rm sh}^4}{8\pi e^4 \ln \Lambda} \nonumber \\
&&\times \frac{1}{4\pi v_{\rm d}^2} \delta \left [v_{\rm d}-\left (1-r^{-1}\right)u_{\rm sh}\right]~~,
\end{eqnarray}
where we assume $u_{\rm rel}\sim u_{\rm sh}$. 
Note that the shock velocity dependence is very strong. 
For leakage neutral particles with a smaller speed of $|v_x|$ 
satisfying $t_{\rm Coulomb}\geq t_{\rm cross,PUI}(v_x)$ 
($v_x\geq -r^{-1}u_{\rm sh}(u_{\rm sh}/u_{\rm sh,c})^4$), 
the distribution function, $f_{\rm leak,sh}(\mbox{\boldmath $v$})$, 
is given by Equation~(\ref{eq:fleaksh3}) because relaxation by Coulomb collision is negligible. 
\section{Relevant Lengthscales}
\label{sec:3}
In this section, we briefly summarize relevant lengthscales in the precursor 
due to leakage neutral particles. 
We here consider charge exchange, collisional ionization and Coulomb collision 
as interactions in the upstream region. 
Ionization of upstream neutral particles by electrons could be important 
compared to ionization by leakage neutral particles and pickup ions 
produced in the upstream region because the number density of electrons 
is larger than that of leakage neutral particles and pickup ions.
However, the electron temperature in the precursor region has not been 
completely understood and ionization by electrons depends on the electron 
temperature. 
Therefore, we here do not take into account ionization by electrons for simplicity.

In the shock rest frame, leakage neutral particles have the mean velocity 
in the direction of the shock normal of $u_{\rm leak,sh}\approx -u_{\rm sh}/3$ 
as shown in Equation~(\ref{eq:uleaksh}).
Then, the relative velocity between upstream particles and leakage neutral 
particles becomes, $u_{\rm rel,leak}\approx 4u_{\rm sh}/3$.
The shock velocity is typically $u_{\rm sh}\sim 3000~{\rm km~s^{-1}}$ 
for young SNRs, so that both collisional ionization with upstream particles 
and charge exchange with upstream ions are important processes 
for leakage neutral particles.
Therefore, the precursor lengthscale in the shock rest frame, $L_{\rm pre}$, 
is given by
\begin{eqnarray}
L_{\rm pre} &=& \frac{|u_{\rm leak,sh}|}{u_{\rm rel,leak}\left \{ \sigma_{\rm i}\left( n_{\rm ion}+n_{\rm n} \right)+\sigma_{\rm ce}n_{\rm ion} \right \}} \nonumber \\
&\approx& 2.5\times10^{15} ~{\rm cm} \left(\frac{|u_{\rm leak,sh}|/u_{\rm rel,leak}}{0.25}\right) \nonumber \\
&&\times \left(\frac{\sigma_{\rm i}+\sigma_{\rm ce}}{10^{-16}~{\rm cm^2}}\right)^{-1}
\left(\frac{n_{\rm ion}+n_{\rm n}}{1~{\rm cm^{-3}}}\right)^{-1}~~,
\end{eqnarray}
where $n_{\rm ion}$ and $n_{\rm n}$ are 
the number densities of the upstream ions and upstream neutral particles, respectively.

Leakage neutral particles are ionized in above region. 
Then, they become pickup ions and are advected into the shock. 
The relaxation lengthscale of pickup ions due to Coulomb collision in the shock rest frame, $L_{\rm Coulomb}$, is given by 
\begin{eqnarray}
L_{\rm Coulomb} &=& \frac{m_{\rm p}^2u_{\rm rel,leak}^3u_{\rm sh}}{8\pi n_{\rm ion}e^4 \ln \Lambda} \nonumber \\
&=& 2\times 10^{21} ~{\rm cm}~\left(\frac{u_{\rm sh}/u_{\rm rel,leak}}{0.75}\right)^{-3} 
\left(\frac{n_{\rm ion}}{0.5~{\rm cm}^{-3}}\right)^{-1} \nonumber \\
&& \times \left(\frac{\ln \Lambda}{40}\right)^{-1}  
\left(\frac{u_{\rm sh}}{3000~{\rm km~s}^{-1}}\right)^{4}~~.
\end{eqnarray}
The relaxation lengthscale, $L_{\rm Coulomb}$, is larger than the precursor 
lengthscale , $L_{\rm pre}$, as long as $u_{\rm sh}>100~{\rm km~s}^{-1}$.
Therefore, Coulomb collision between pickup ions and upstream ions 
is negligible, that is, pickup ions do not relax to upstream ions 
in the precursor region as long as $u_{\rm sh}>100~{\rm km~s}^{-1}$.

Another loss process of pickup ions in the precursor region is charge exchange 
between pickup ions and neutral particles. 
The charge exchange lengthscale
of pickup ions in the shock rest frame, $L_{\rm ce,PUI}$, 
is given by
\begin{eqnarray}
L_{\rm ce,PUI} &=& \frac{u_{\rm sh}}{u_{\rm rel,leak}\sigma_{\rm ce}(n_{\rm n}+n_{\rm leak,sh})} \nonumber \\
&=& 1.5\times 10^{16}~{\rm cm}~\left(\frac{u_{\rm sh}/u_{\rm rel,leak}}{0.75}\right)  \nonumber \\
&&\times \left(\frac{\sigma_{\rm ce}}{10^{-16}~{\rm cm}^2}\right)^{-1} \left(\frac{n_{\rm n}+n_{\rm leak,sh}}{0.5~{\rm cm}^{-3}}\right)^{-1}~~.
\label{eq:lcepui}
\end{eqnarray}
The precursor lengthscale , $L_{\rm pre}$, is always smaller than 
the charge exchange lengthscale of pickup ions, $L_{\rm ce,PUI}$, 
as long as the leakage velocity, $|u_{\rm leak,sh}|$, is smaller than 
the shock velocity, $u_{\rm sh}$.
Therefore, we can neglect charge exchange of pickup ions 
for $|u_{\rm leak,sh}| < u_{\rm sh}$.

Upstream neutral particles are ionized by collision and charge exchange 
with leakage neutral particles and pickup ions.
The ionization lengthscale of upstream neutral particles in the shock rest frame, $L_{\rm i,up}$,  is given by
\begin{eqnarray}
L_{\rm i,up}&=& \frac{u_{\rm sh}}{u_{\rm rel,leak}\left \{ \sigma_{\rm i}(n_{\rm leak,sh}+n_{\rm PUI})+\sigma_{\rm ce}n_{\rm PUI} \right \} } \nonumber \\
&\approx& 1.5\times10^{17} ~{\rm cm} \left(\frac{u_{\rm sh}/u_{\rm rel,leak}}{0.75}\right) \nonumber \\
&&\times \left(\frac{\sigma_{\rm i}+\sigma_{\rm ce}}{10^{-16}~{\rm cm^2}}\right)^{-1}
\left(\frac{n_{\rm leak,sh}+n_{\rm PUI}}{0.05~{\rm cm^{-3}}}\right)^{-1}~~,
\end{eqnarray}
where $n_{\rm PUI}$ is the number density of pickup ions and 
we assume $n_{\rm leak,sh}\sim 0.1n_{\rm n}\sim 0.05~{\rm cm}^{-3}$ 
as shown in Equation (\ref{eq:nleaksh}).
Therefore, the ionization of upstream neutral particles is negligible 
in the precursor because $L_{\rm i,up}>L_{\rm pre}$.

Upstream neutral particles interact not only with leakage neutral particles 
but also with upstream ions.
The relative velocity between upstream neutral particles and upstream 
ions, $u_{\rm rel,up}$, would become larger than their thermal velocity but 
smaller than the shock velocity ($\sim 3000~{\rm km~s}^{-1}$) 
for high Mach number shocks. 
Therefore, charge exchange is dominant and 
the charge exchange lengthscale of upstream neutral particles with 
upstream ions in the shock rest frame, $L_{\rm ce,up}$, is given by
\begin{eqnarray}
L_{\rm ce,up}&=&\frac{u_{\rm sh}}{u_{\rm rel,up}\sigma_{\rm ce}n_{\rm ion}} \nonumber \\
&=&2\times10^{16}~{\rm cm} \left(\frac{u_{\rm sh}/u_{\rm rel,up}}{10}\right) \nonumber \\
&&\times \left(\frac{\sigma_{\rm ce}}{10^{-15}~{\rm cm^2}}\right)^{-1}
\left(\frac{n_{\rm ion}}{0.5~{\rm cm^{-3}}}\right)^{-1}~~.
\end{eqnarray}
where we assume that upstream ions are decelerated to 
about 90 percent of the shock velocity, $0.9 u_{\rm sh}$ (see Section \ref{sec:4}).
If leakage is small, the upstream plasma flow does not change 
significantly, so that $L_{\rm ce,up}>L_{\rm pre}$.
Then, upstream neutral particles rarely interact with upstream ions 
in the precursor, that is, the flow velocity of upstream neutral particles 
does not change.
We consider this decoupling case in Section~\ref{sec:4.1}.
If leakage is large or the shock velocity is larger than $3000~{\rm km~s}^{-1}$, 
the upstream plasma flow is significantly decelerated, 
$u_{\rm rel,up}\sim u_{\rm sh}$, or 
the precursor lengthscale becomes large because the ionization cross section 
becomes small, so that $L_{\rm ce,up}<L_{\rm pre}$. 
Then, upstream neutral particles interact many times with upstream ions 
in the precursor.
As the result, the velocity distribution of upstream neutral particles quickly 
becomes that of upstream ions at each point in the precursor, that is, 
the flow velocity (mean velocity) of upstream neutral particles becomes 
that of upstream ions.
However, it should be noted that the relative velocity between upstream ions 
and upstream neutral particles is not approximately zero 
because of their thermal velocity.
We consider this tight-coupling case in Section~\ref{sec:4.2}.

According to the DSA theory, accelerated particles diffuse into 
the upstream region.
The diffusion lengthscale is given by
\begin{eqnarray}
L_{\rm diff} &=& \frac{\eta_{\rm g}cE}{3eBu_{\rm sh}} \nonumber \\
&=& 10^{15}~{\rm cm} \left(\frac{\eta_{\rm g}}{1}\right) \left(\frac{u_{\rm sh}}{3000~{\rm km~s^{-1}}}\right)^{-1} \nonumber \\
&&\times \left(\frac{B}{100~{\rm \mu G}}\right)^{-1}
\left(\frac{E}{1~{\rm TeV}}\right)~~,
\end{eqnarray}
where $\eta_{\rm g}, B$ and $E$ are the gyrofactor, the magnetic field 
and the energy of accelerated particles.
Therefore, particles are accelerated in the precursor due to leakage 
neutral particles, that is, leakage of neutral particles is important for 
the particle acceleration up to $1$-$10~{\rm TeV}$.
The diffusion lengthscale of pickup ions is much smaller than the precursor scale, $L_{\rm pre}$.
Hence, we can neglect diffusion of pickup ions.

\section{Four Fluid Model in a Neutral Particle Precursor}
\label{sec:4}
In this section, we calculate the steady-state precursor structure due to 
leakage neutral particles. 
There are cold ions and neutral particles in the upstream region.
In addition, we consider leakage neutral particles and pickup ions originating 
from leakage neutral particles.
We here adopt a four fluids model to describe the precursor. 
Our treatment is similar to that of \citet{hengetal07}. 
Continuity equations of the steady state are given by
\begin{eqnarray}
\label{eq:con_n}
\frac{\rm d}{{\rm d} x} \left(n_{\rm n} u_{\rm n} \right)&=& - Q_{\rm ion}~~,  \\
\frac{\rm d}{{\rm d} x} \left(n_{\rm leak} u_{\rm leak} \right)  &=& - Q_{\rm PUI}~~, \\
\frac{\rm d}{{\rm d} x} \left(n_{\rm ion} u_{\rm ion} \right)&=&    Q_{\rm ion}~~,  \\
\label{eq:con_pui}
\frac{\rm d}{{\rm d} x} \left(n_{\rm PUI} u_{\rm ion} \right)&=&    Q_{\rm PUI}~~,
\end{eqnarray}
where the subscripts 'n', 'leak', 'ion' and 'PUI' denote upstream neutral particles, 
leakage neutral particles, upstream ions and pickup ions, respectively and 
$x$ is the coordinate of the direction along the shock normal 
($x=0$ and $x=-\infty$ are the positions of the shock and the far upstream region,
 respectively.). 
We assume that the fluid velocity of pickup ions is the same as that of 
upstream ions because of electromagnetic interactions, $u_{\rm PUI}=u_{\rm ion}$.
$Q_{\rm ion}$ and $Q_{\rm PUI}$ are source terms and given by 
\begin{eqnarray} 
Q_{\rm ion}  &=& (n_{\rm leak} + n_{\rm PUI}) n_{\rm n} \sigma_{\rm i} u_{\rm rel,leak} \nonumber \\
&&-n_{\rm ion}n_{\rm leak}\sigma_{\rm ce}u_{\rm rel,leak}~~,\\
Q_{\rm PUI} &=& (n_{\rm n}+n_{\rm PUI}+n_{\rm ion}) n_{\rm leak} \sigma_{\rm i} u_{\rm rel,leak} \nonumber \\
&&+n_{\rm ion}n_{\rm leak}\sigma_{\rm ce}u_{\rm rel,leak}~~,
\end{eqnarray}
where we approximate all relative velocities as a constant 
(see Equation~(\ref{eq:urel})).
The first two terms of $Q_{\rm ion}$ are due to 
collisional ionization of upstream neutral particles with 
leakage neutral particles and pickup ions, respectively.
The last term of $Q_{\rm ion}$ is due to charge exchange of 
upstream ions with leakage neutral particles. 
The first three terms of $Q_{\rm PUI}$ are due to 
collisional ionization of leakage neutral particles with upstream 
neutral particles, pickup ions and upstream ions, respectively. 
The last term of $Q_{\rm PUI}$ is due to charge exchange of 
leakage neutral particles with upstream ions. 
For the decoupling case ($L_{\rm ce,up}>L_{\rm pre}$), 
we can neglect charge exchange and collisional ionization 
between upstream ions and upstream neutral particles 
because their interaction lengthscales are larger than 
the precursor lengthscale.
For the tight-coupling case ($L_{\rm ce,up}<L_{\rm pre}$), 
we approximately treat upstream ions and neutral particles 
as a single fluid at each point in the precursor 
instead for solving charge exchange processes between upstream ions 
and upstream neutral particles. 
Once we assume the tight-coupling between upstream ions and 
upstream neutral particles, 
charge exchange between upstream ions 
and neutral particles approximately does not change anything 
and collisional ionization between upstream ions and neutral 
particles does not change dynamics of the single fluid. 
Therefore, we do not take into account charge exchange 
and collisional ionization between upstream ions and neutral particles 
for both cases. 
Moreover, we can neglect charge exchange of pickup ions as discussed 
in Equation~(\ref{eq:lcepui})

In the next subsections, we solve Equations (\ref{eq:con_n})-(\ref{eq:con_pui}) 
by using momentum and energy conservations.
Boundary conditions are as follows:
\begin{eqnarray}
n_{\rm n}(-\infty)&=&n_{\rm n,0}~~~~~,~u_{\rm n}(-\infty)=u_0 ~~, \\
n_{\rm ion}(-\infty)&=&n_{\rm ion,0}~~~,~u_{\rm ion}(-\infty)=u_0 ~~, \\
n_{\rm PUI}(-\infty)&=&0 ~~, \\
n_{\rm leak}(0)&=&n_{\rm leak,sh}~,u_{\rm leak}(0)=u_{\rm leak,sh} ~~,
\end{eqnarray}
where the subscripts '0' and 'sh' represent quantities 
at the far upstream region and at the shock, respectively.
Note that $u_{\rm leak,sh}$ is negative.
All quantities are normalized by $u_0$ and $n_{\rm ion,0}+n_{\rm n,0}$ 
in Sections \ref{sec:4.1} and \ref{sec:4.2}.
Although we estimated $n_{\rm leak,sh}\approx 0.1n_{\rm n,0}$ and $u_{\rm leak,sh}\approx -u_0/3$ 
in Section~\ref{sec:3}, there are a few uncertainties, 
especially for the velocity distribution of downstream hot ions. 
Hence, we treat $n_{\rm n,0}/(n_{\rm ion,0}+n_{\rm n,0}), 
n_{\rm leak,sh}/(n_{\rm ion,0}+n_{\rm n,0})$ and $u_{\rm leak,sh}/u_0$ 
as free parameters in Sections~\ref{sec:4.1} and \ref{sec:4.2}.

We treat leakage neutral particles as a cold fluid in this paper, that is, 
we neglect the velocity dispersion of leakage neutral particles.
Then, the relative velocity, $u_{\rm rel,leak}$, is approximately given by  
\begin{equation}
u_{\rm rel,leak} = u_0 - u_{\rm leak,sh}~~.
\label{eq:urel}
\end{equation}
%
\subsection{Decoupling Approximation ($L_{\rm ce,up}>L_{\rm pre}$)}
\label{sec:4.1}
\begin{figure}
\plotone{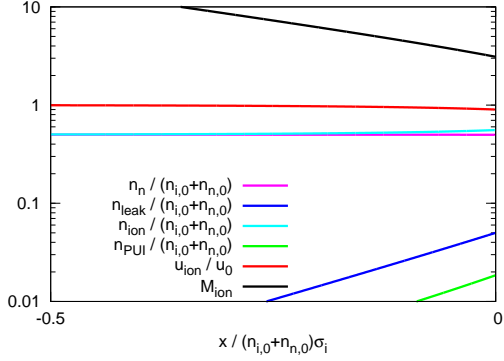} 
\caption{Number densities and velocity profiles for the decoupling approximation. 
The magenta, blue, cyan and green lines show number densities of upstream 
neutral particles, leakage neutral particles, upstream ions and pickup ions, 
respectively. The red and black lines show the ion flow velocity and the Mach 
number, respectively. Boundary conditions are 
$M_{\rm ion,0} = 100, n_{\rm n,0}/(n_{\rm ion,0}+n_{\rm n,0})=0.5, 
n_{\rm leak,sh}/(n_{\rm ion,0}+n_{\rm n,0})=0.05$ 
and $u_{\rm leak,sh}/u_0=-1/3$. 
The other parameter is $\sigma_{\rm ce}/\sigma_{\rm i}=1$. 
\label{fig:1}}
\end{figure}
\begin{figure}
\plotone{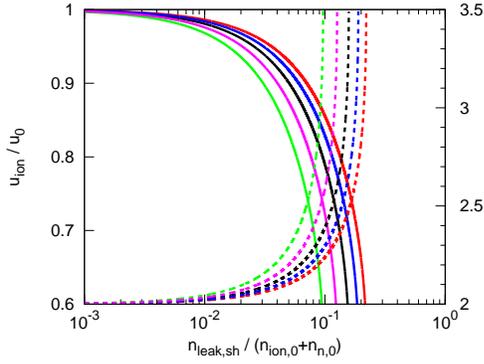} 
\caption{The fluid velocity of upstream ions at the shock (solid lines) and 
the spectral index of accelerated particles (dashed lines) 
for the decoupling approximation. 
The black, red, green, blue and magenta lines show cases of $\left(n_{\rm n,0}/(n_{\rm ion,0}+n_{\rm n,0}), u_{\rm leak,sh}/u_0 \right)= (0.5, -1/3),  (0.5, -1/4), (0.5, -1/2), (0.4, -1/3)$ and $(0.6, -1/3)$, respectively.
The other boundary condition is $M_0=M_{\rm ion,0}=100$. 
\label{fig:2}}
\end{figure}

In this subsection, we solve equations of the four fluid system using the 
decoupling approximation corresponding to small leakage of neutral particles.
We here assume that fluid velocities of upstream neutral particle and leakage 
neutral particles do not change from that at the far upstream region, $u_0$, 
and that at the shock, $u_{\rm leak,sh}$, respectively. 
Then, continuity equations are given by
\begin{eqnarray}
\label{eq:con_n2}
u_0\frac{\rm d}{{\rm d} x} n_{\rm n} &=& - Q_{\rm ion}~~,\\
\label{eq:con_leak2}
u_{\rm leak,sh}\frac{\rm d}{{\rm d} x} n_{\rm leak}  &=& - Q_{\rm PUI}~~,\\
\label{eq:con_ion2}
\frac{\rm d}{{\rm d} x} \left(n_{\rm ion} u_{\rm ion} \right)&=&    Q_{\rm ion}~~,\\
\label{eq:con_pui2}
\frac{\rm d}{{\rm d} x} \left(n_{\rm PUI} u_{\rm ion} \right)&=&    Q_{\rm PUI}~~.
\end{eqnarray}
Momentum and energy conservation laws of 
upstream ions and pickup ions are approximately given by
\begin{eqnarray}
\frac{\rm d}{{\rm d} x} \left\{m\left(n_{\rm ion}+n_{\rm PUI}\right)u_{\rm ion}^2 + P_{\rm ion} +P_{\rm PUI}\right\} \nonumber \\
= m\left(u_0 Q_{\rm ion}  + u_{\rm leak,sh}Q_{\rm PUI}\right)~~,
\label{eq:mcon1}
\end{eqnarray}
\begin{eqnarray}
\frac{\rm d}{{\rm d} x} \left\{\frac{1}{2}m(n_{\rm ion}+n_{\rm PUI})u_{\rm ion}^3
+ \frac{\gamma}{\gamma-1}u_{\rm ion} \left( P_{\rm ion}+P_{\rm PUI}\right) \right\} \nonumber \\
=\frac{1}{2}m\left(u_0^2 Q_{\rm ion}+u_{\rm leak,sh}^2Q_{\rm PUI}\right)~~,
\label{eq:encon1}
\end{eqnarray}
where $m, P_{\rm ion},  P_{\rm PUI}$ and $\gamma$ are 
the particle mass, the pressure of upstream ions, the pressure of pickup ions 
and the adiabatic index, respectively.
In the first term of the right hand side of 
Equations~(\ref{eq:mcon1}) and (\ref{eq:encon1}), 
we approximate the fluid velocity of upstream ions as that at far upstream region, 
$u_{\rm ion}\approx u_0$. 
This approximation is valid as long as deceleration of upstream ions is small.

To make the expression simple, hereafter velocities, densities, pressures 
and spatial coordinate are normalized by $u_0, (n_{\rm ion,0}+n_{\rm n,0}), 
m(n_{\rm ion,0}+n_{\rm n,0})u_0^2$ and 
$(n_{\rm ion,0}+n_{\rm n,0})^{-1}\sigma_{\rm i}^{-1}$, respectively.
Normalized quantities are denoted with a bar.
From Equations~(\ref{eq:con_n2})-(\ref{eq:encon1}), 
one can make four conserved quantities.
Hence, Equations~(\ref{eq:con_n2})-(\ref{eq:encon1}) reduce to 
the following two equations: 
\begin{eqnarray}
\label{eq:con_n3}
\frac{{\rm d}}{{\rm d} {\bar x}}  {\bar n_{\rm n}}&=&
-{\bar u_{\rm rel,leak}}{\bar n_{\rm leak}}
\left[{\bar n_{\rm n}}- \frac{1}{{\bar u_{\rm ion}}} \right. \nonumber \\
&&\times \left. \left \{ \left(1-{\bar n_{\rm n}} \right) \left(\frac{\sigma_{\rm ce}}{\sigma_{\rm i}}\right) + {\bar n_{\rm n}} {\bar u_{\rm leak,sh}} \right \} \right] ~~,\\
\label{eq:con_leak3}
\frac{{\rm d}}{{\rm d} {\bar x}}  {\bar n_{\rm leak}}&=&
-\frac{{\bar u_{\rm rel,leak}}}{{\bar u_{\rm leak,sh}}} {\bar n_{\rm leak}} \left[{\bar n_{\rm n}} +\frac{1}{{\bar u_{\rm ion}}} \right. \nonumber \\
&&\times \left. \left\{\left(1-{\bar n_{\rm n}}\right)\left(1+\frac{\sigma_{\rm ce}}{\sigma_{\rm i}}\right)-{\bar n_{\rm leak}}{\bar u_{\rm leak,sh}} \right\}\right]~~,
\end{eqnarray}
where ${\bar u_{\rm ion}}$ can be expressed by
\begin{eqnarray}
\label{eq:u_ion}
{\bar u_{\rm ion}}&=&\frac{B+\sqrt{B^2-4AC}}{2A}~~,\\
A&=& \left(\gamma+1\right)\left(1-{\bar n_{\rm n}}-{\bar n_{\rm leak}}{\bar u_{\rm leak,sh}}\right)~~, \nonumber \\
B &=&2\gamma\left(1+\frac{{\bar n_{\rm ion,0}}}{\gamma M_{\rm ion,0}^2}-{\bar n_{\rm n}}-{\bar n_{\rm leak}}{\bar u_{\rm leak,sh}}^2\right)~~,\nonumber \\
C &=&(\gamma-1)\left(1+\frac{2{\bar n_{\rm ion,0}}}{\left(\gamma -1\right)M_{\rm ion,0}^2}-{\bar n_{\rm n}}-{\bar n_{\rm leak}}{\bar u_{\rm leak,sh}}^3\right)\nonumber ~~,
\end{eqnarray}
where $M_{\rm ion,0}=({\bar n_{\rm ion,0}}/\gamma {\bar P_{\rm ion,0}})^{1/2}$ is the Mach number of ions at the far upstream region.
From Equations~(\ref{eq:con_n3})-(\ref{eq:u_ion}), 
one can obtain solutions of ${\bar n_{\rm n}}({\bar x}), 
{\bar n_{\rm leak}}({\bar x})$ and ${\bar u_{\rm ion}}({\bar x})$ 
by numerical computations.
By assuming that the brackets term of Equation~(\ref{eq:con_leak3}) 
is constant, one can obtain analytical approximations which are valid 
for ${\bar n_{\rm leak,sh}} \ll1$.
Using the solutions of ${\bar n_{\rm n}}, {\bar n_{\rm leak}}$ 
and ${\bar u_{\rm ion}}$, the other quantities, ${\bar n_{\rm ion}}, 
{\bar n_{\rm PUI}}$ and ${\bar P_{\rm ion}}+{\bar P_{\rm PUI}}$ 
can be expressed by
\begin{eqnarray}
{\bar n_{\rm ion}} &=& \frac{1-{\bar n_{\rm n}}}{{\bar u_{\rm ion}}}~~,\\
{\bar n_{\rm PUI}}&=& -\frac{{\bar u_{\rm leak,sh}}}{{\bar u_{\rm ion}}}{\bar n_{\rm leak}}~~,\\
{\bar P_{\rm ion}}+{\bar P_{\rm PUI}}&=& 1+\frac{{\bar n_{\rm ion,0}}}{\gamma M_{\rm ion,0}^2}-{\bar n_{\rm n}}-{\bar n_{\rm leak}}{\bar u_{\rm leak,sh}}^2 \nonumber \\
&&-\left(1-{\bar n_{\rm n}}-{\bar n_{\rm leak}}{\bar u_{\rm leak,sh}}\right){\bar u_{\rm ion}}~~.
\end{eqnarray}
Moreover, the evolution of the Mach number of upstream ions and pickup ions, 
$M_{\rm ion}({\bar x})$, can be expressed by
\begin{eqnarray}
M_{\rm ion} &=& {\bar u_{\rm ion}}\sqrt{ \frac{ {\bar n_{\rm ion}} + {\bar n_{\rm PUI}}}{\gamma \left( {\bar P_{\rm ion}}+{\bar P_{\rm PUI}} \right)} }\nonumber \\
&=& \gamma^{-\frac{1}{2}} \left\{ \frac{1+\frac{{\bar n_{\rm ion,0}}}{\gamma M_{\rm ion,0}^2}-{\bar n_{\rm n}}-{\bar n_{\rm leak}}{\bar u_{\rm leak,sh}}^2}{\left(1-{\bar n_{\rm n}}-{\bar n_{\rm leak}}{\bar u_{\rm leak,sh}}\right){\bar u_{\rm ion}}}-1\right\}^{-\frac{1}{2}}~~.
\label{eq:mach_de}
\end{eqnarray}

Figure~\ref{fig:1} shows numerical solutions to 
Equations ({\ref{eq:con_n3}) - ({\ref{eq:mach_de}), 
where $M_{\rm ion,0}=100, \gamma=5/3, {\bar n_{\rm n,0}}=0.5, 
{\bar n_{\rm leak,sh}}=0.05, {\bar u_{\rm leak,sh}}=-1/3$ 
and $\sigma_{\rm ce}/\sigma_{\rm i}=1$.
For the same input parameters, analytical approximations 
of Equation (\ref{eq:con_leak3}) give about $0.1$ percent accuracy. 
There is no solution with $M_{\rm ion}<1$.
The flow velocity of upstream ions and pickup ions, ${\bar u_{\rm ion}}$, is slightly 
decelerated by small leakage of neutral particles. 
However, the Mach number becomes small significantly because the pressure 
of pickup ions is large.
These features are qualitatively the same as results of \citet{blasi12}.
The number density of upstream neutral particles, 
${\bar n_{\rm n}}$, is almost constant.
As already mentioned in Section~\ref{sec:3}, 
ionization of upstream neutral particles is negligible 
in the precursor region for ${\bar n_{\rm leak,sh}} \ll 1$.
Therefore, the number density of upstream neutral particles, ${\bar n_{\rm n}}$, 
can be regarded as a constant.
Then, one can obtain analytical solutions at the shock.

Because we have all quantities at the shock, 
we can calculate the shock jump condition.
Collisionless shocks are formed only by the plasma because the 
dissipation length of the plasma is much smaller than the interaction length 
of neutral particles \citep{chevalier78,chevalier80}.
It should be noted that we have to take into account sink terms due to 
leakage of neutral particles when we derive the Rankin-Hugoniot relations 
at the shock. 
Even though leakage neutral particles are not ions, the origin is hot ions in the 
downstream region. 
By using the Rankin-Hugoniot relations between the far upstream region and the 
downstream region, one can easily obtain the shock jump condition. 
This is because there is no net sink between the far upstream region 
and the downstream region. 
Therefore, the total compression ratio between the far upstream region and the 
downstream region is given by 
\begin{equation}
r_{\rm tot} = \frac{\gamma+1}{\gamma-1+2M_0^{-2}} ~~,
\end{equation}
where $M_0=\left \{ \left({\bar n_{\rm ion,0}}+{\bar n_{\rm n,0}} \right)/\gamma \left({\bar P_{\rm ion,0}} + {\bar P_{\rm n,0}} \right) \right \}^{1/2}=M_{\rm ion,0}$ 
is the Mach number of all particles at the far upstream region. 
The partial compression ratio between the far upstream region 
and the front of the shock (${\bar x}=-\epsilon$) is ${\bar u_{\rm ion}}^{-1}$, 
so that the compression ratio of the shock is given by
\begin{equation}
r_{\rm sh} = \frac{\gamma+1}{\gamma-1+2M_0^{-2}} {\bar u_{\rm ion}}({\bar x}=0)~~.
\label{eq:rsh}
\end{equation}
One can obtain the analytical approximation of the shock compression ratio, 
$r_{\rm sh}$, from Equations~(\ref{eq:u_ion}) and (\ref{eq:rsh}) 
because of ${\bar n_{n}}\approx {\bar n_{\rm n,0}}$.
Figure~\ref{fig:2} shows the analytical approximations of the fluid velocity of upstream 
ions at the shock, ${\bar u_{\rm ion}}({\bar x}=0)$, and the spectral index of 
accelerated particles, $s=(r_{\rm sh}+2)/(r_{\rm sh}-1)$. 
Even though leakage is small (${\bar n_{\rm leak,sh}}\lesssim 0.1$), 
the spectral index becomes larger than 2. 
As already mentioned by \citet{blasi12},
this can explain the observed gamma-ray spectra slightly steeper 
than the simplest prediction of DSA.
Effects of leakage neutral particles become significant 
for the low ionization fraction and the large leakage flux.
Note that the decoupling approximation may not be valid for 
${\bar n_{\rm leak,sh}}\gtrsim0.1$ because the relative velocity between 
upstream neutral particles and upstream ions becomes large (see Section 3).
\subsection{Tight-Coupling Approximation ($L_{\rm ce,up}<L_{\rm pre}$)}
\label{sec:4.2}
\begin{figure}
\plotone{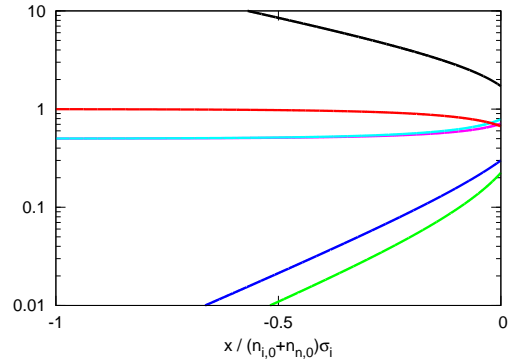} 
\caption{The same as Figure~\ref{fig:1}, 
but for the tight-decoupling approximation. 
Boundary conditions are $M_0 = 100, n_{\rm n,0}/(n_{\rm ion,0}+n_{\rm n,0})=0.5, 
n_{\rm leak,sh}/(n_{\rm ion,0}+n_{\rm n,0})=0.3$ and $u_{\rm leak,sh}/u_0=-0.5$.
The other parameter is $\sigma_{\rm ce}/\sigma_{\rm i}=1$. 
\label{fig:3}}
\end{figure}
\begin{figure}
\plotone{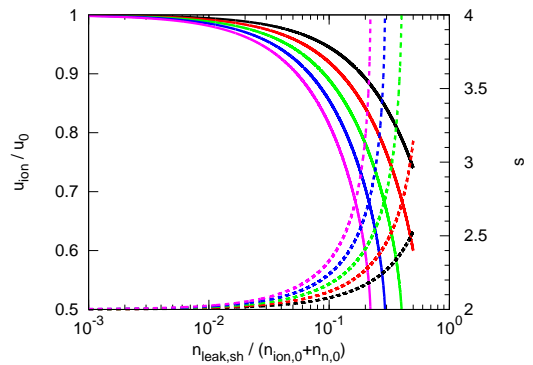} 
\caption{The same as Figure~\ref{fig:2}, 
but for the tight-decoupling approximation. 
The black, red, green, blue and magenta lines show cases of 
$u_{\rm leak,sh}/u_0= -0.3, -0.4, -0.5, -0.6$ and $-0.7$, respectively.
The other boundary condition is $M_0=100$. 
\label{fig:4}}
\end{figure}

In this subsection, we solve equations of the four fluid system using the 
tight-coupling approximation corresponding to large leakage of neutral particles.
We here assume that the fluid velocity of leakage neutral particles does 
not change from that at the shock, $u_{\rm leak,sh}$. 
In addition, we assume that the fluid velocity of upstream neutral particles 
is the same as that of upstream ions because of the tight coupling, 
$u_{\rm n}=u_{\rm ion}$.
Then, continuity equations are given by
\begin{eqnarray}
\label{eq:con_n1t}
\frac{\rm d}{{\rm d} x} \left(n_{\rm n} u_{\rm ion} \right)&=& - Q_{\rm ion} ~~, \\
u_{\rm leak,sh}\frac{\rm d}{{\rm d} x} \left(n_{\rm leak}  \right)  &=& - Q_{\rm PUI}~~, \\
\frac{\rm d}{{\rm d} x} \left(n_{\rm ion} u_{\rm ion} \right)&=&Q_{\rm ion}~~,  \\
\frac{\rm d}{{\rm d} x} \left(n_{\rm PUI} u_{\rm ion} \right)&=&Q_{\rm PUI}~~.
\end{eqnarray}
Momentum and energy conservation laws of upstream neutral particles, 
upstream ions and pickup ions are given by
\begin{eqnarray}
\frac{\rm d}{{\rm d} x} \left\{m\left(n_{\rm n}+n_{\rm ion}+n_{\rm PUI}\right)u_{\rm ion}^2 + P_{\rm n}+P_{\rm ion} +P_{\rm PUI}\right\} \nonumber \\
=  mu_{\rm leak,sh}Q_{\rm PUI}~~,
\end{eqnarray}
\begin{eqnarray}
&&\frac{\rm d}{{\rm d} x} \left\{\frac{1}{2}m(n_{\rm n}+n_{\rm ion}+n_{\rm PUI})u_{\rm ion}^3 \right. \nonumber \\
&&~~~~~~~~~~~~~~~\left. + \frac{\gamma}{\gamma-1}u_{\rm ion} \left( P_{\rm n}+P_{\rm ion}+P_{\rm PUI}\right) \right\} \nonumber \\
&&=\frac{1}{2}mu_{\rm leak,sh}^2Q_{\rm PUI}~~.
\label{eq:energy_t}
\end{eqnarray}
where $P_{\rm n}$ is the pressure of upstream neutral particles.

To make the expression simple, hereafter velocities, densities, pressures 
and spatial coordinate are normalized by $u_0, (n_{\rm ion,0}+n_{\rm n,0}), 
m(n_{\rm ion,0}+n_{\rm n,0})u_0^2$ and 
$(n_{\rm ion,0}+n_{\rm n,0})^{-1}\sigma_{\rm i}^{-1}$, respectively.
Normalized quantities are denoted with a bar.
From Equations~(\ref{eq:con_n1t})-(\ref{eq:energy_t}), 
one can make four conserved quantities.
Hence, Equations~(\ref{eq:con_n1t})-(\ref{eq:energy_t}) reduce to 
the following two equations: 
\begin{eqnarray}
\label{eq:con_nt2}
\frac{{\rm d}}{{\rm d} {\bar x}}  \left( {\bar n_{\rm n}{\bar u_{\rm ion}}}\right)&=&
-{\bar u_{\rm rel,leak}}{\bar n_{\rm leak}}
\left[{\bar n_{\rm n}} \left(1+\frac{\sigma_{\rm ce}}{\sigma_{\rm i}}\right) \right. \nonumber \\
&&\left. -\frac{1}{{\bar u_{\rm ion}}}  \left \{ {\bar n_{\rm n}}{\bar u_{\rm leak,sh}} +\left(\frac{\sigma_{\rm ce}}{\sigma_{\rm i}} \right) \right \} \right] ~~,\\
\label{eq:con_leakt}
\frac{{\rm d}}{{\rm d} {\bar x}}  {\bar n_{\rm leak}}&=&
-\frac{ {\bar u_{\rm rel,leak}}}{{\bar u_{\rm leak,sh}}} {\bar n_{\rm leak}}
\left[ -\left(\frac{\sigma_{\rm ce}}{\sigma_{\rm i}}\right) {\bar n_{\rm n}}\right. \nonumber \\
&&+\left. \frac{1}{{\bar u_{\rm ion}}} \left\{1+\left(\frac{\sigma_{\rm ce}}{\sigma_{\rm i}}\right)-{\bar n_{\rm leak}} {\bar u_{\rm leak,sh}} \right\} \right]~~,
\end{eqnarray}
where ${\bar u_{\rm ion}}$ can be expressed by
\begin{eqnarray}
\label{eq:uiont}
{\bar u_{\rm ion}}&=&\frac{B+\sqrt{B^2-4AC}}{2A}~~,\\
A&=& \left(\gamma+1\right)\left(1-{\bar n_{\rm leak}}{\bar u_{\rm leak,sh}}\right)~~, \nonumber \\
B &=&2\gamma\left(1+\frac{1}{\gamma M_0^2}-{\bar n_{\rm leak}}{\bar u_{\rm leak,sh}}^2\right)~~,\nonumber \\
C &=&(\gamma-1)\left(1+\frac{2}{\left(\gamma -1\right)M_0^2}-{\bar n_{\rm leak}}{\bar u_{\rm leak,sh}}^3\right)\nonumber ~~,
\end{eqnarray}
From Equations~(\ref{eq:con_nt2})-(\ref{eq:uiont}), one can obtain solutions of 
${\bar n_{\rm n}}({\bar x}), {\bar n_{\rm leak}}({\bar x})$ and 
${\bar u_{\rm ion}}({\bar x})$ by numerical computations.
By assuming that the brackets terms of Equations (\ref{eq:con_nt2}) and (\ref{eq:con_leakt}) are constant, one can obtain analytical approximations.
Using the solutions of ${\bar n_{\rm n}}, {\bar n_{\rm leak}}$ and ${\bar u_{\rm ion}}$, 
the other quantities, ${\bar n_{\rm ion}}, {\bar n_{\rm PUI}}$ and 
${\bar P_{\rm n}}+{\bar P_{\rm ion}}+{\bar P_{\rm PUI}}$ can be expressed by
\begin{eqnarray}
{\bar n_{\rm ion}} &=& \frac{1}{{\bar u_{\rm ion}}} -{\bar n_{\rm n}}~~,\\
{\bar n_{\rm PUI}}&=& -\frac{{\bar u_{\rm leak,sh}}}{{\bar u_{\rm ion}}}{\bar n_{\rm leak}}~~,\\
{\bar P_{\rm n}}+{\bar P_{\rm ion}}+{\bar P_{\rm PUI}}&=& 1+\frac{1}{\gamma M_0^2}-{\bar n_{\rm leak}}{\bar u_{\rm leak,sh}}^2 \nonumber \\
&&-\left(1-{\bar n_{\rm leak}}{\bar u_{\rm leak,sh}}\right){\bar u_{\rm ion}}~~.
\end{eqnarray}
Moreover, the evolution of the Mach number, $M({\bar x})$,  can be expressed by
\begin{eqnarray}
M &=& {\bar u_{\rm ion}}\sqrt{ \frac{ {\bar n_{\rm n}}+ {\bar n_{\rm ion}} + {\bar n_{\rm PUI}}}{\gamma \left( {\bar P_{\rm n}}+{\bar P_{\rm ion}}+{\bar P_{\rm PUI}} \right)} }\nonumber \\
&=& \gamma^{-\frac{1}{2}} \left\{ \frac{1+\frac{1}{\gamma M_0^2}-{\bar n_{\rm leak}}{\bar u_{\rm leak,sh}}^2}{\left(1-{\bar n_{\rm leak}}{\bar u_{\rm leak,sh}}\right){\bar u_{\rm ion}}}-1\right\}^{-\frac{1}{2}}~~.
\label{eq:macht}
\end{eqnarray}

Figure~\ref{fig:3} shows numerical solutions to Equations ({\ref{eq:con_nt2}) - 
({\ref{eq:macht}), where $M_0=100, \gamma=5/3, {\bar n_{\rm n,0}}=0.5, 
{\bar n_{\rm leak,sh}}=0.3$, ${\bar u_{\rm leak,sh}}=-0.5$.
For the same input parameters, analytical approximations of Equations 
(\ref{eq:con_nt2}) and (\ref{eq:con_leakt}) give an about $5$ percent accuracy. 
There is no solution with $M<1$.
Unlike the decoupling approximation, the flow velocity of upstream neutral 
particles, upstream ions and pickup ions, ${\bar u_{\rm ion}}$, is significantly 
decelerated by large leakage of neutral particles. 
The number densities of upstream neutral particles, 
${\bar n_{\rm n}}$, and upstream ions, ${\bar n_{\rm ion}}$, 
become large because of compression.
Moreover, the increase rate of the number density of upstream ions is slightly larger than
that of upstream neutral particles because some of upstream neutral particles 
are ionized in the precursor region.
Unlike the decoupling approximation, the flow velocity of upstream ions and 
the Mach number, Equations (\ref{eq:uiont}) and (\ref{eq:macht}), do not explicitly 
depend on the number density of upstream neutral particles.
Therefore, one can obtain analytical solutions at the shock.
From Equation (\ref{eq:rsh}), one can obtain the analytical solution of the shock 
compression ratio, $r_{\rm sh}$.
Figure~\ref{fig:4} shows the analytical solutions of the fluid velocity 
of upstream ions at the shock, ${\bar u_{\rm ion}}({\bar x}=0)$, 
and the spectral index of accelerated particles, $s=(r_{\rm sh}+2)/(r_{\rm sh}-1)$. 
For ${\bar n_{\rm leak,sh}}\sim 0.1$, the spectral index becomes larger than 2. 
As already mentioned by \citet{blasi12},
this can explain the observed gamma-ray spectra slightly steeper 
than the simplest prediction of DSA.
Note that the tight-coupling approximation is not valid for 
${\bar n_{\rm leak,sh}}\ll 0.1$ because the fluid velocity of upstream ions 
does not  significantly change (see Section 3).
\section{Discussion}
We here discuss other important effects of leakage neutral particles.
When leakage neutral particles are ionized, their velocity distribution
in the upstream rest frame is initially a beam-like or ring-like distribution.
These velocity distributions excite electromagnetic fields and amplify the magnetic 
field \citep{wu72,lee87,raymond08,ohiraetal09,ohira10}. 
This is a promising mechanism to explain some observations concerning to 
the strong magnetic field \citep{vink03,berezhko03,bamba05,uchiyama07}.
Moreover, the electromagnetic instabilities could heat the upstream region.
H$\alpha$ observed from the upstream region \citep{lee10} can be interpreted 
as the results of leakage neutral particles.

In Sections~\ref{sec:4.1} and \ref{sec:4.2}, we assumed that the adiabatic 
index of pickup ions is $5/3$.
However, there is no guarantee that the behavior of pickup ions
is the same as the standard gas because of the collisionless system.
Especially, the behavior of pickup ions at the shock is important 
for the shock jump condition and particle accelerations.
Even though the Mach number at the shock is not so small, 
the compression ratio could be smaller than 4
because the adiabatic index of pickup ions could be larger than $5/3$ 
\citep{fahr08,wu09}.
If pickup ions drain the large fraction of the shock kinetic energy, 
upstream ions are not heated up to $T=3mu_{\rm sh}^2/16$.
This can explain the recent observation of H$\alpha$ \citep{helder09}
which showed that the temperature derived from the line width of 
H$\alpha$ is much smaller than derived from the proper motion.
Furthermore, pickup ions produced in the precursor region could be 
preferentially accelerated by DSA because their velocity is larger than 
that of upstream ions \citep{ohira10}. 
These issues are crucial not only for particle accelerations and 
shock dissipation but also for the amount of leakage neutral particles from the 
downstream region to the upstream region.
Therefore, we treated values concerning to the leakage neutral particles as 
free parameters in Sections~\ref{sec:4.1} and \ref{sec:4.2}.

In this paper, we did not specify neutral particles.
Helium atoms have a smaller cross section than that of hydrogen atoms.
We expect large lengthscale of the precursor region compared with that 
of hydrogen atoms.
However, the ionization fraction of helium depends on time because 
helium atoms are ionized by radiation from the downstream region \citep{gha00}.
Therefore, the precursor lengthscale and the injection of helium ions into DSA 
could depend on an SNR age.
The CR injection history of helium ions is important to understand 
the CR helium spectrum observed at the Earth \citep{drury11,ohira11}.
\section{Summary}
In this paper, we have investigated effects of leakage neutral particles on shocks.
We have found that if the dominant source of leakage neutral particles is 
pickup ions produced in the downstream region, 
about ten percent of upstream neutral particles leak into 
the upstream region (Equation~(\ref{eq:nleaksh})) and 
the mean leakage velocity is about one third of the shock velocity 
(Equation~(\ref{eq:uleaksh})). 
Moreover, we have calculated the precursor structure due to 
the leakage neutral particles by using four fluid approximations (upstream neutral particles, 
upstream ions, leakage neutral particles and pickup ions). 
We have found analytical solutions of the precursor structure 
by using the decoupling approximation or the tight-coupling approximation, 
where the decoupling means that upstream 
neutral particles do not interact with upstream ions and the tight-coupling
means the opposite case.
We have found that even when leakage is small, the shock compression ratio 
becomes significantly small. 
This can explain the observed gamma-ray spectra slightly steeper 
than the simplest prediction of DSA.
In addition, leakage neutral particles could heat the precursor region and 
pickup ions produced in the precursor region are important for the injection into DSA.

\acknowledgments
We thank T. Terasawa and S. Matsukiyo for useful comments about physics of the solar wind. 
We also thank the anonymous referee for valuable comments. 
This work is supported in part by grant-in-aid from the Ministry of Education, 
Culture, Sports, Science, and Technology (MEXT) of Japan, No.~24$\cdot$8344.


\begin{thebibliography}{}
%
\bibitem[Abdo et al.(2009)]{abdo09}
Abdo, A. A. et al., 2009, \apj, 706, L1
%
\bibitem[Abdo et al.(2010)]{abdo10}
Abdo, A. A., et al., 2010, ApJ, 710, L92
%
\bibitem[Axford et al.(1977)]{axford77}
Axford, W. I., Leer, E., \& Skadron, G., 1977, Proc. 15th Int. Cosmic Ray Conf., Plovdiv, 11, 132
%
\bibitem[Bamba et al.(2005)]{bamba05}
Bamba, A., Yamazaki R., Yoshida T., Terasawa T., and Koyama, K., 2005, \apj, 621, 793
%
\bibitem[Barnett et al.(1990)]{barnett90}
Barnett, C. F., Hunter, H. T., Kirkpatrick, M.I., Alvarez, I., Cisneros, C., \& Phaneuf, R. A., 1990, Atomic Data for Fusion, Vol. 1, ORNL-6086/V1
%
\bibitem[Blandford \& Ostriker(1978)]{blandford78}
Blandford, R. D., \& Ostriker, J. P., 1978, \apj, 221, L29
%
\bibitem[Bell(1978)]{bell78}
Bell, A. R., 1978, \mnras, 182, 147
%
\bibitem[Bell et al.(2012)]{bell12}
Bell, A. R., Schure, K. M., Reville, B., 2012, \mnras, 418, 1208 
%
\bibitem[Berezhko et al.(2003)]{berezhko03} Berezhko, E. G., Ksenofontov, L. T., \& V{\"o}lk, H. J., 2003 \aap, 412, L11
%
\bibitem[Blasi et al.(2012)]{blasi12}Blasi, P., Morlino, G., Bandiera, R., Amato, E., Caprioli, D., 2012, \apj, 755, 121
%
\bibitem[Cassam-Chena\"{i} et al.(2008)]{cassam08}
Cassam-Chena\"{i}, G., Hughes, J. P., Reynoso, E. M., Badenes, C., \& Moffett, D., 2008, \apj, 680, 1180
%
\bibitem[Caprioli et al.(2010)]{caprioli10}
Caprioli, D., Amato, E., \& Blasi, P., 2010, Astropart. Phys., 33, 160
%
\bibitem[Chevalier \& Raymond(1978)]{chevalier78} Chevalier, R. A. \& Raymond, J. C., 1978, \apj, 225, L27
%
\bibitem[Chevalier et al.(1980)]{chevalier80} Chevalier, R. A., Kirshner, R. P., \& Raymond, J. C., 1980, \apj, 235, 186
%
\bibitem[Drury et al.(1996)]{drury96}Drury, L. O'C., Duffy, P., \& Kirk, J. G., 1996, \aap, 309, 1002
%
\bibitem[Drury(2011)]{drury11}Drury, L. O'C., 2011, \mnras, 415, 1807
%
\bibitem[Fahr \& Chalov(2008)]{fahr08}Fahr, H. J., \& Chalov, S. V. 2008, \aap, 409, L35
%
\bibitem[Ghavamian et al.(2000)]{gha00}
Ghavamian, Raymond, J., Hartigan, P., \& Blair, W. P., 2000, \apj, 535, 266
%
\bibitem[Ghavamian et al.(2002)]{gha02}
Ghavamian, Winkler, P. F., Raymond, J. C., \& Long, K. S., 2002, \apj, 572, 888
%
\bibitem[Giordano et al.(2012)]{giordano12}Giordano, F., et al. 2012, \apj, 744, L2
%
\bibitem[Giuliani et al.(2011)]{giuliani11}
Giuliani, A., et al. 2011, \apj, 742, L30
%
\bibitem[Gloeckler et al.(1993)]{gloeckler93}
Gloeckler, G., et al. 1993, Science, 261, 70
%
\bibitem[Helder et al.(2009)]{helder09}
Helder, E. A., Vink, J., Bassa, C. G., Bamba, A., Bleeker, J. A. M., Funk, S., Ghavamian, P., van der Heyden, K. J., Verbunt, F., and Yamazaki, R., 2009, Science, 325, 719
%
\bibitem[Heng \& McCray(2007)]{heng07}
Heng, K., \& McCray, R., 2007, \apj, 654, 923
%
\bibitem[Heng et al.(2007)]{hengetal07}
Heng, K., van Adelsberg, M., McCray, R., \& Raymond, J. C., 2007, \apj, 668, 275
%
\bibitem[Heng(2010)]{heng10}
Heng, K., 2010, PASA, 27, 23
%
\bibitem[Kirk et al.(1996)]{kirk96}
Kirk J. G., Duffy P., Gallant Y. A., 1996, \aap, 314, 1010
%
\bibitem[Krymsky (1977)]{krymsky77}
Krymsky, G. F., 1977, Dokl. Akad. Nauk SSSR, 234, 1306
%
\bibitem[Lee \& Ip(1987)]{lee87}Lee, M., \& Ip, W.-H. 1987, \jgr, 92, 11041
%
\bibitem[Lee et al.(2010)]{lee10}
Lee, J.-J., Raymond, J. C., Park, S., Blair, W. P., Ghavamian, P., Winkler, P. F., \& Korreck, K., 2010, \apj, 715, L146
%
\bibitem[McComas et al.(2009)]{mccomas09}
McComas, D. J., et al., 2009, Science, 326, 959
%
\bibitem[Ohira \& Takahara(2007)]{ohira07}
Ohira, Y., \& Takahara, F., 2007, \apj, 661, 171
%
\bibitem[Ohira \& Takahara(2008)]{ohira08}
Ohira, Y., \& Takahara, F., 2008, \apj, 688, 320
%
\bibitem[Ohira et al.(2009)]{ohiraetal09}
Ohira, Y., Terasawa, T., \& Takahara, F., 2009, \apj, 703, L59
%
\bibitem[Ohira \& Takahara(2010)]{ohira10}
Ohira, Y., \& Takahara, F., 2010, \apjl, 721, L43
%
\bibitem[Ohira et al.(2010)]{ohiraetal10}
Ohira, Y., Murase, K. \& Yamazaki, R., 2010, \aap, 513, A17
%
\bibitem[Ohira et al.(2011)]{ohiraetal11}
Ohira, Y., Murase, K. \& Yamazaki, R., 2011, \mnras, 410, 1577
%
\bibitem[Ohira \& Ioka(2011)]{ohira11}
Ohira, Y., \& Ioka, K., 2011, \apj, 729, L13
%
\bibitem[Ptuskin \& Zirakashvili(2005)]{ptuskin05}
Ptuskin, V. S., \& Zirakashvili, V. N., 2005, \aap, 429, 755
%
\bibitem[Rakowski et al.(2008)]{rakowski08}
Rakowski, C. E., Laming, J. M., \& Ghavamian, P., 2008, \apj, 684, 348 
%
\bibitem[Raymond et al.(2008)]{raymond08}Raymond, J. C., Isenberg. P. A., \& Laming, J. M., 2008, \apj, 682, 408 
%
\bibitem[Reville et al.(2007)]{reville07}Reville, B., Kirk. J. G., Duffy, P., \& O'Sullivan, 2007, \aap, 475, 435 
%
\bibitem[Spitzer(1962)]{spitzer62} Spitzer, L., 1962, Physics of Fully Ionized Gases (2nd ed,; New York: Wiley)
%
\bibitem[Tavani et al.(2010)]{tavani10}
Tavani, M. et al., 2010, \apj, 710, L151
%
\bibitem[Uchiyama et al.(2007)]{uchiyama07}
Uchiyama, Y., Aharonian, F A., Tanaka, T., Takahashi, T., \& Maeda, Y., 2007, \nat, 449, 576
%
\bibitem[Vink \& Laming(2003)]{vink03}
Vink, J., \& Laming, J. M., 2003, \apj, 584, 758
%
\bibitem[Wu \& Davidson(1972)]{wu72}Wu, S. C., \& Davidson, R. C., 1972, \jgr, 77, 5399 
%
\bibitem[Wu et al.(2009)]{wu09}
Wu, P., Winske, D., Gary, S. P., Schwadron, N. A. \& Lee, M. A., 2009, \jgr, 114, A08103 
%
\bibitem[Zirakashvili \& Ptuskin(2009)]{zirakashvili09}
Zirakashvili V. N., Ptuskin V. S., 2009, in Aharonian F. A., Hofmann W.,
Rieger F. M. eds, AIP Conf. Proc. Vol. 1085, High Energy Gamma-Ray
Astronomy, Am. Inst. Phys., New York, p. 336
%
\end{thebibliography}
\end{document}